\documentclass[preprint,12pt]{elsarticle}

\usepackage[T1]{fontenc}
\usepackage{lmodern}
\usepackage{placeins}

\usepackage{amsmath,amssymb}
\usepackage{graphicx}
\usepackage{subcaption}
\usepackage{float}
\usepackage{booktabs}
\usepackage{siunitx}
\usepackage{xcolor}

\journal{Physica A: Statistical Mechanics and its Applications}

\begin{document}

\begin{frontmatter}

\title{Geometry-controlled clogging and transport mechanisms in confined Y-junctions}


\author[unsj]{Horacio Javier Diaz Carrera}
\author[ib,inn]{D. J. Perez-Morelo}
\author[unsl,INFAP]{Sergio J. Manzi}
\author[unsl,unsj,INFAP]{Paulo M. Centres\corref{cor1}}
\ead{paulo.centres@unsl.edu.ar}
\cortext[cor1]{Corresponding author}


\affiliation[unsl]{
	organization={Departamento de F\'isica,
		Universidad Nacional de San Luis},
	city={San Luis},
	country={Argentina}
}

\affiliation[unsj]{
	organization={Departamento de Geof\'isica,
		Astronom\'ia y F\'isica,
		Universidad Nacional de San Juan},
	addressline={Mitre 396 (E)},
	postcode={J5402CWH},
	city={San Juan},
	country={Argentina}
}


\affiliation[INFAP]{
	organization={Instituto de F\'isica Aplicada
		``Dr. Jorge Andr\'es Zgrablich'' (INFAP),
		CONICET--Universidad Nacional de San Luis},
	addressline={Av. Ej\'ercito de los Andes 950},
	postcode={5700},
	city={San Luis},
	country={Argentina}
}

\affiliation[ib]{
	organization={Instituto Balseiro,
		Universidad Nacional de Cuyo},
	city={San Carlos de Bariloche},
	country={Argentina}
}

\affiliation[inn]{
	organization={Consejo Nacional de Investigaciones
		Cient\'ificas y T\'ecnicas (CONICET),
		INN--CNEA},
	city={San Carlos de Bariloche},
	country={Argentina}
}

\begin{abstract}
	We investigate transport and clogging of hard particles in a confined symmetric Y-junction using off-lattice Monte Carlo simulations. The geometry is controlled by the channel width $w$ and a truncation parameter $s$ that modifies the inner vertex. The flow probability exhibits a sharp sigmoidal crossover, defining a characteristic width $w_c$ that depends non-monotonically on $s$. This non-monotonic behavior persists under variations of the transverse Monte Carlo step, although the absolute values of $w_c$ change. Microscopic analysis reveals two distinct blocking mechanisms: single-particle obstruction at small truncations and collective arch-like structures at larger truncations. Residence-time statistics and particle-order inversions provide additional evidence of this change in transport dynamics. Our results demonstrate that local modifications of junction geometry can strongly affect clogging and particle transport, even in a minimal model with excluded-volume interactions only.
\end{abstract}


\end{frontmatter}

\section{Introduction}
\label{sec:introduction}

Single-file diffusion (SFD) describes particle transport under strong geometrical confinement, where particles cannot overtake one another and their relative order is preserved. This constraint leads to anomalous tagged-particle dynamics, characterized by subdiffusive behavior at long times \cite{Hodgkin,Taloni,Harris,Jepsen,Kasperkovitz,Arriata,Richards,Fedders,Alexander,Lizana,Beijeren}. A similar preservation of particle order occurs in ballistic single-file systems (BSF), despite their different microscopic dynamics \cite{Coste,Delfau,Tkachenko,Centres3}.

The relaxation of geometrical confinement introduces additional degrees of freedom, allowing particles to exchange positions and modifying their transport properties. This behavior has been investigated in coupled lattice channels \cite{Kutner} and continuum systems of confined hard particles \cite{Lucena,Mon,Jimaan,Ooshida,Centres4}, where particle-passing events lead to a crossover from single-file behavior to ordinary diffusion. These studies demonstrate how geometrical constraints govern particle motion and collective dynamics.

An abrupt change in confinement, as occurs in channel junctions, introduces a different scenario. Local geometrical features such as Y-junctions \cite{Larimi,Antonov2025,Bushi2005}, T-junctions \cite{Garstecki2006}, and serpentine microchannels \cite{Zhang2014} may induce particle rearrangements, modify their initial ordering, and generate local congestion. Under sufficiently strong confinement, these rearrangements may also produce transient or persistent flow arrest.

Flow arrest is closely related to clogging phenomena observed in granular systems, where stable arch-like configurations can obstruct particle transport through narrow openings \cite{Zuriguel2005,To2001,To2002,To2005}. In a Y-junction, however, particles encounter an extended region connected to two competing branches, where different blocking configurations may emerge depending on the local geometry.

Beyond flow arrest, the junction introduces waiting times associated with particle interactions and local congestion. The statistical characterization of these delays connects confined particle transport with broader problems involving residence and waiting times, such as traffic flow and queueing processes. In this context, both the preservation of particle order and the distribution of residence times provide complementary information about the transport mechanisms.

Here, we investigate the dynamics of hard disks moving through a symmetric Y-junction using off-lattice Monte Carlo simulations. The particles interact exclusively through excluded volume, while a longitudinal bias drives them toward the bifurcation. We independently vary the channel width $w$ and a truncation parameter $s$ controlling the shape of the inner vertex. We show that the characteristic width $w_c$ exhibits a non-monotonic dependence on $s$, associated with distinct microscopic blocking mechanisms. This geometrical dependence is accompanied by changes in particle-order inversions and residence-time statistics, revealing how local relaxation of confinement can profoundly modify collective transport. The non-monotonic behavior remains robust under variations of the transverse Monte Carlo step.

\section{Model and simulation method}
\label{sec:model}

\subsection{Geometry}

We consider $N=36$ identical hard disks of diameter $\phi$ confined to a two-dimensional symmetric Y-junction. The main channel has length $L\simeq23.2\phi$, while each outgoing branch has length $L/2$ and forms an angle $\alpha=30^{\circ}$ with the longitudinal axis. All lengths are expressed in units of $\phi$.

\begin{figure}[htbp]
	\centering
	\includegraphics[
	width=0.95\linewidth,
	keepaspectratio
	]{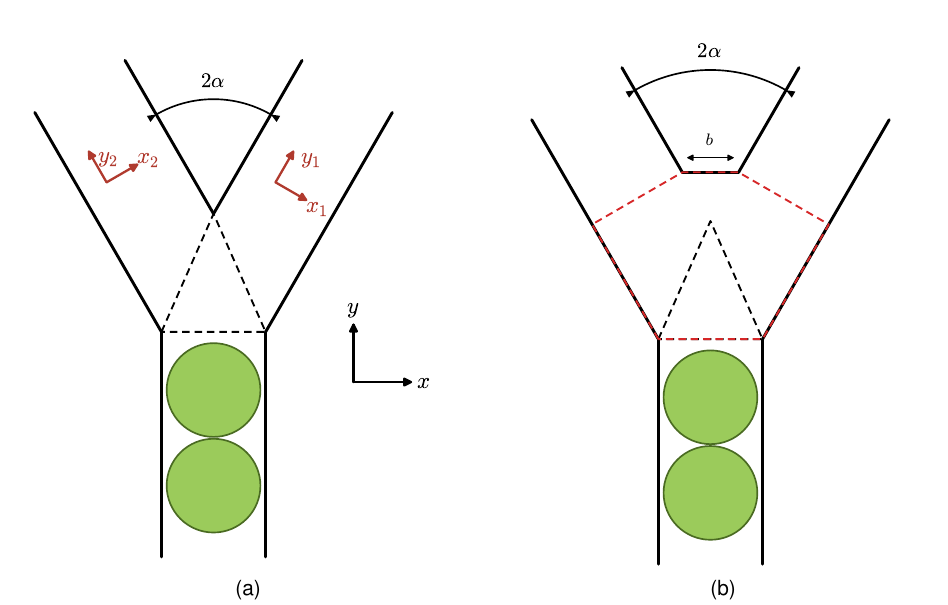}
	
	\caption{Geometry of the symmetric Y-junction:
		(a) ideal bifurcation ($s=0$);
		(b) truncated bifurcation ($s>0$), with a flat
		inner vertex of length $b=s\phi$.
		The dashed contours illustrate the change
		in the bifurcation region, whose area increases by
		$\Delta A_{\mathrm{bif}}(s)=
		A_{\mathrm{bif}}(s)-A_{\mathrm{bif}}(0)$.}
	
	\label{fig:model_geometry}
\end{figure}

The channel width is $W=w\phi$, with $1<w<2$, preventing particle overtaking in the straight sections and preserving their relative order. The inner vertex is characterized by a dimensionless truncation parameter $s$. For $s=0$, the inner walls meet at an ideal point [Fig.~\ref{fig:model_geometry}(a)], whereas for $s>0$, the vertex is replaced by a flat segment of length $b=s\phi$ [Fig.~\ref{fig:model_geometry}(b)]. Thus, $w$ controls the degree of confinement, while $s$ modifies the local geometry of the bifurcation without changing the channel width or branching angle.

\begin{figure}[htbp]
	\centering
	\includegraphics[
	height=6.5cm,
	keepaspectratio
	]{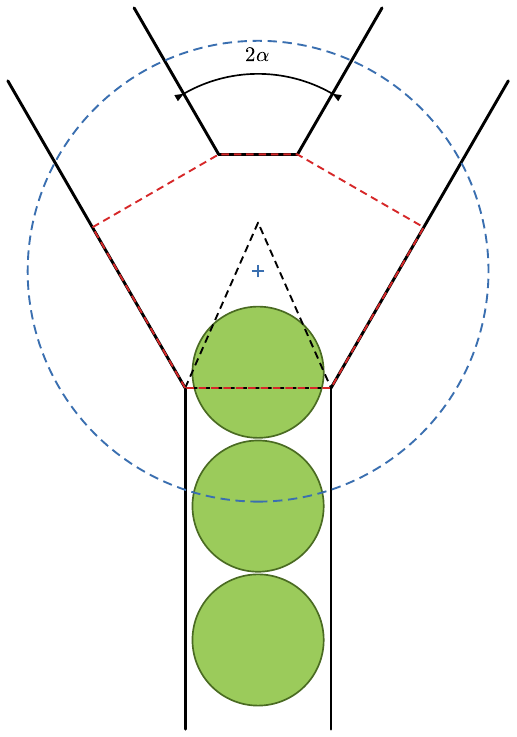}
\caption{Circular detection region used to measure
	particle residence times near the Y-junction.
	The dashed circle represents the entry boundary
	of radius $R_{\mathrm{det}}$.
	To avoid spurious boundary crossings caused by
	particle fluctuations, exit is recorded only
	when the particle center crosses an outer
	boundary of radius $R_{\mathrm{det}}+0.5\phi$,
	not shown for clarity. This spatial hysteresis
	prevents artificial fragmentation of the
	residence-time intervals.}
\label{fig:detection_region}
\end{figure}

Particle transit is characterized using a circular detection region centered on the bifurcation [Fig.~\ref{fig:detection_region}]. The residence time is defined as the interval between particle entry and exit. To avoid spurious crossings caused by fluctuations near the boundary, entry is recorded when the particle center crosses a circle of radius $R_{\mathrm{det}}$, whereas exit requires crossing an outer circle of radius $R_{\mathrm{det}}+0.5\phi$. This spatial hysteresis prevents artificial fragmentation of residence-time measurements.

\subsection{Monte Carlo dynamics}
\label{sec:MC}

Particle dynamics is simulated using off-lattice Monte Carlo moves. In the main channel, transverse displacements are proposed with probabilities $P_{\mathrm{left}}=P_{\mathrm{right}}=0.25$, while longitudinal moves have probabilities $P_+=0.26$ and $P_-=0.24$, introducing a weak bias toward the junction. The corresponding step lengths are
\begin{equation}
	\Delta y=2.5\times10^{-4}\phi,
	\qquad
	\Delta x=\frac{(w-1)\phi}{n_T},
\end{equation}
with $n_T=2000$ in the main simulations. The transverse step scales with the available lateral space, yielding an acceptance rate close to $0.99$ over the explored widths. Additional simulations with $n_T=3000$ confirm that the non-monotonic dependence of the characteristic width $w_c$ on the truncation parameter $s$ is preserved, although its absolute values depend on the transverse step.

Moves are accepted only when particle overlaps and wall crossings are avoided. One Monte Carlo step (MCS) consists of one attempted move per active particle. Each realization spans $T_{\mathrm{obs}}=2\times10^7$ MCS. Particles exiting through either branch are reinserted at the main-channel entrance when sufficient space becomes available, maintaining a constant total particle number.

Clogging episodes are identified using a binary indicator $\chi(t)$ based on particle accumulation, structural persistence, and local mobility within the detection region. Let $A(t)$ be the set of particles inside this region, $n(t)=|A(t)|$ its occupancy, and
$q(t)=|A(t)\cap A(t-1)|/|A(t)\cup A(t-1)|$
the Jaccard overlap. The indicator is defined as
\begin{equation}
	\chi(t)=
	\begin{cases}
		1, & n(t)\geq4,\quad q(t)\geq0.80,\\
		& \langle\Delta r^2(t)\rangle\leq(0.005\phi)^2,\\
		0, & \text{otherwise},
	\end{cases}
\end{equation}
where $\langle\Delta r^2(t)\rangle$ is evaluated over particles present at consecutive MCS. Changes in the clogging state are confirmed after 100 consecutive MCS satisfying the corresponding condition. An unclogging event is accepted only if a particle subsequently crosses the junction; otherwise, adjacent episodes are merged.

The occupancy threshold distinguishes clogging from ordinary transport, typically involving three particles in the detection region. Persistent blockages involving fewer than four particles, which may escape detection by $\chi(t)$, are identified independently through the absence of completed crossings during the observation window.

\subsection{Flow probability and critical width}

For each realization $m$, the flow fraction is defined as $p_m=N_m^*/N$, where $N_m^*$ is the number of particles completing at least one transit through the junction during the observation time $T_{\mathrm{obs}}$. Averaging over $M=75$ independent realizations gives
\begin{equation}
	P_f(w,s)=\frac{1}{M}\sum_{m=1}^{M}p_m.
\end{equation}

For fixed $s$, the flow probability is fitted by a logistic function,
\begin{equation}
	P_f(w)=\frac{1}{1+\exp[(w_c-w)/\delta w]},
	\label{eq:logistic}
\end{equation}
where $w_c$ is the characteristic width defined by $P_f(w_c)=1/2$, and $\delta w$ characterizes the crossover width. Since $w_c$ is determined over a finite observation time, it represents an operational rather than a strictly geometric threshold.

\subsection{Residence times and order inversions}

For each completed transit $i$ through the detection region defined in Fig.~\ref{fig:detection_region}, the residence time is
\begin{equation}
	\tau_i=t_i^{\mathrm{out}}-t_i^{\mathrm{in}}.
\end{equation}
The corresponding survival function is defined as
\begin{equation}
	S(\tau)=P(T>\tau),
\end{equation}
where $T$ denotes the residence time of a particle.

Since particle overtaking is forbidden in the straight channels, changes in particle order can occur only within the junction region. We quantify these rearrangements through $f_{\mathrm{inv}}$, defined as the fraction of consecutive particle pairs whose exit order differs from their entrance order.

\section{Results and discussion}
\label{sec:results}

\subsection{Geometry-dependent flow transition}

\begin{figure}[htbp]
	\centering
	\includegraphics[width=0.48\linewidth]{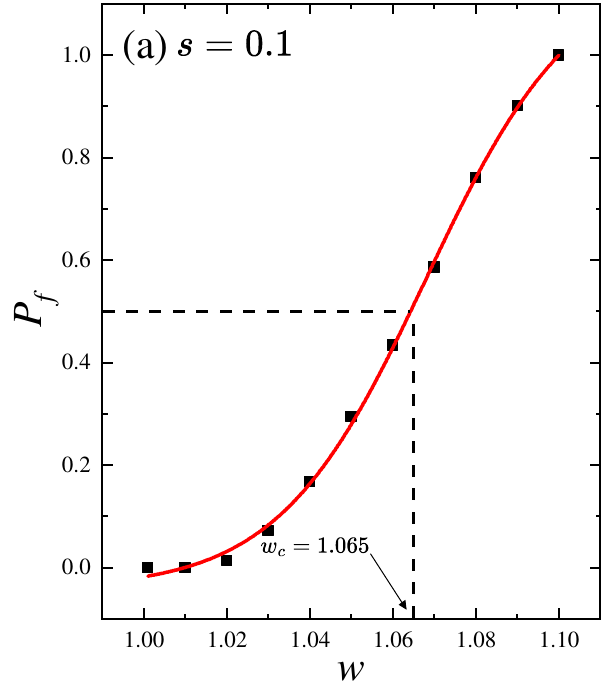}
	\hfill
	\includegraphics[width=0.48\linewidth]{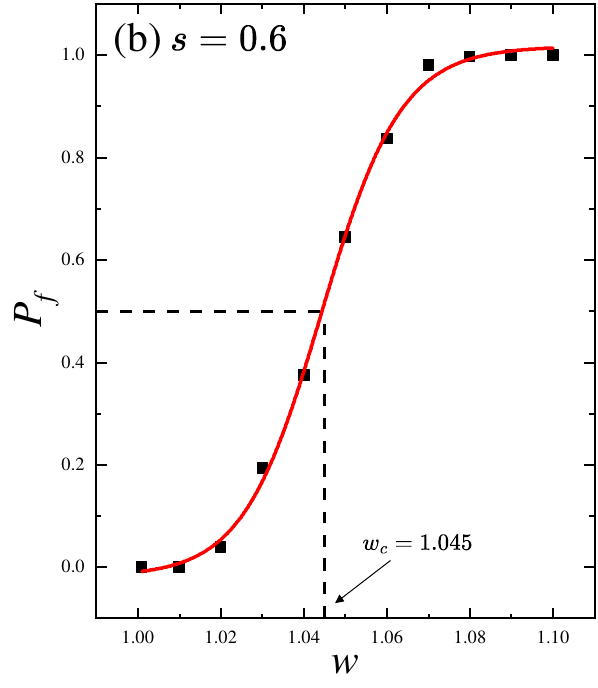}
	
	\caption{Flow probability $P_f$ versus channel
		width $w$ for (a) $s=0.1$ and (b) $s=0.6$.
		Symbols represent simulation results,
		solid lines are logistic fits, and dashed
		lines indicate the characteristic widths
		$w_c$, defined by $P_f(w_c)=1/2$.}
	
	\label{fig:flow_probability}
\end{figure}

For all truncation parameters $s$, the flow probability $P_f$ exhibits a sharp sigmoidal increase with channel width $w$, marking a crossover from predominantly blocked to flowing configurations. Figure~\ref{fig:flow_probability} shows representative results for $s=0.1$ and $s=0.6$, together with the logistic fits defined by Eq.~\eqref{eq:logistic}.

\begin{figure}[htbp]
	\centering
	\includegraphics[width=0.75\linewidth]{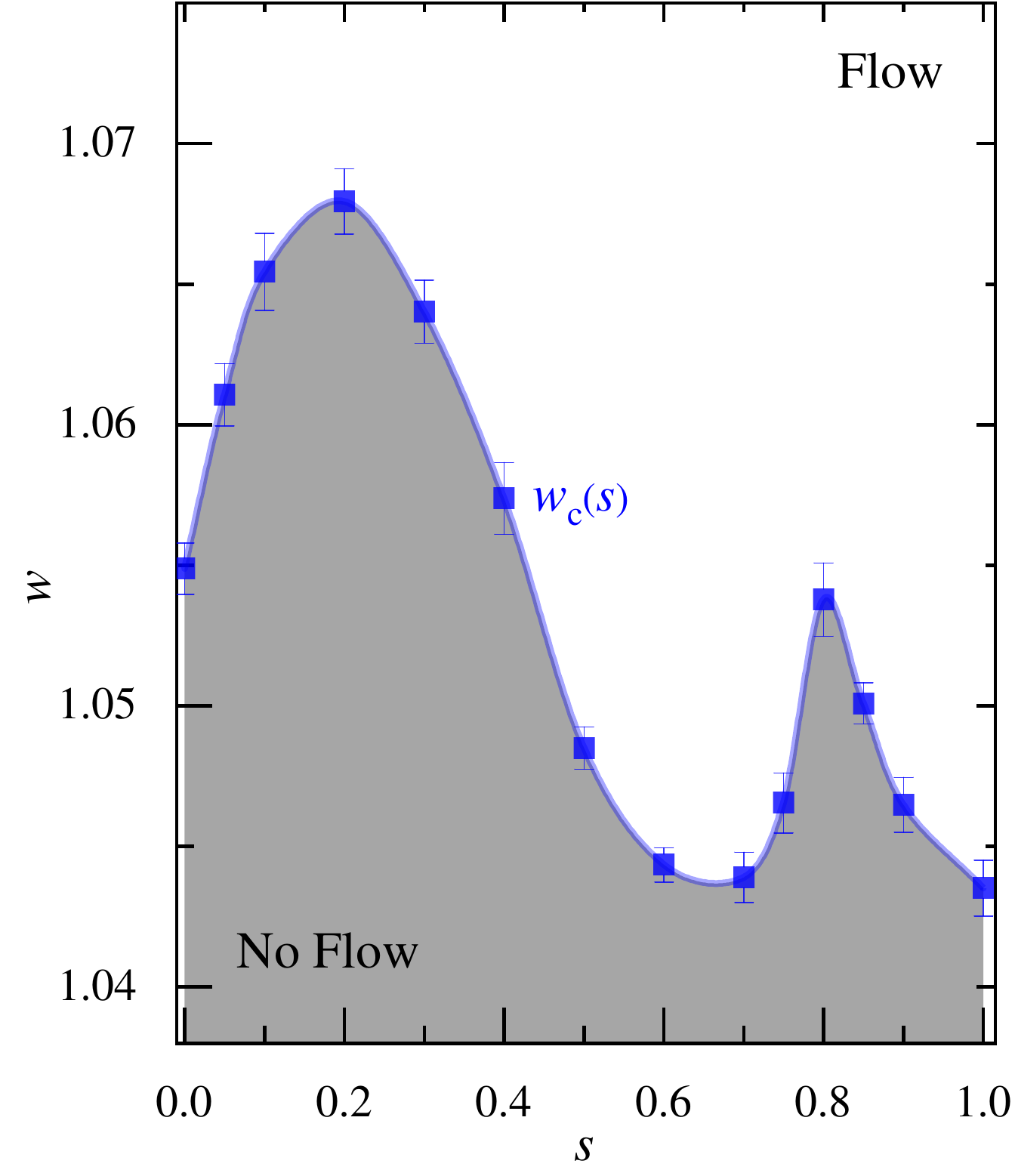}
	\caption{Characteristic width $w_c$ as a function
		of the truncation parameter $s$. Symbols
		represent values obtained from logistic fits,
		with error bars indicating fitting uncertainties.
		The curve separates predominantly blocked
		and flowing regimes within the finite
		observation time.}
	\label{fig:phase_diagram}
\end{figure}

The main result is the non-monotonic dependence of $w_c$ on the local junction geometry. As shown in Fig.~\ref{fig:phase_diagram} and Table~\ref{tab:wc}, $w_c$ reaches a global maximum of $1.0679$ at $s=0.2$, decreases to $1.0439$ at $s=0.7$, and exhibits a secondary maximum at $s=0.8$. These variations demonstrate that local changes in the junction geometry strongly affect transport, even when the channel width and branching angle remain unchanged.

\begin{table}[H]
	\centering
	\caption{Characteristic widths $w_c$ obtained from logistic fits to the flow probability. Uncertainties correspond to fitting errors.}
	\label{tab:wc}
	\begin{tabular}{cc@{\qquad}cc}
		\toprule
		$s$ & $w_c$ & $s$ & $w_c$ \\
		\midrule
		0.000 & $1.0549\pm0.0009$ & 0.600 & $1.0444\pm0.0006$ \\
		0.050 & $1.0611\pm0.0011$ & 0.700 & $1.0439\pm0.0009$ \\
		0.100 & $1.0654\pm0.0014$ & 0.750 & $1.0466\pm0.0011$ \\
		0.200 & $1.0679\pm0.0012$ & 0.800 & $1.0538\pm0.0013$ \\
		0.300 & $1.0640\pm0.0011$ & 0.850 & $1.0501\pm0.0007$ \\
		0.400 & $1.0574\pm0.0013$ & 0.900 & $1.0465\pm0.0010$ \\
		0.500 & $1.0485\pm0.0008$ & 1.000 & $1.0435\pm0.0010$ \\
		\bottomrule
	\end{tabular}
\end{table}

The sensitivity of $w_c(s)$ to the transverse Monte Carlo
step and the observation time is examined in Appendix~A.
Although the absolute values of $w_c$ depend on these
simulation parameters, the non-monotonic dependence
on $s$ and its main features are preserved over the
investigated parameter ranges.

The characteristic widths obtained here are smaller than those typically reported for clogging in granular hoppers and silos \cite{To2005,Hong2017}. However, this comparison is qualitative, since the physical mechanisms and definitions of flow probability differ. In particular, $w_c$ is determined from finite-duration simulations and therefore represents an operational crossover width rather than a strict critical opening. The results establish the influence of local geometry on transport within the investigated dynamical and temporal conditions, without implying the existence of an absolute clogging threshold.

\subsection{Microscopic origin of the non-monotonic threshold}

To investigate the microscopic origin of the non-monotonic dependence of $w_c(s)$, we analyzed particle configurations near the inner vertex for five representative geometries ($s=0$, $0.2$, $0.7$, $0.8$, and $1.0$), considering six channel widths between $w=1.030$ and $1.080$.

\begin{figure}[!htbp]
	\centering
	\includegraphics[width=0.95\textwidth]{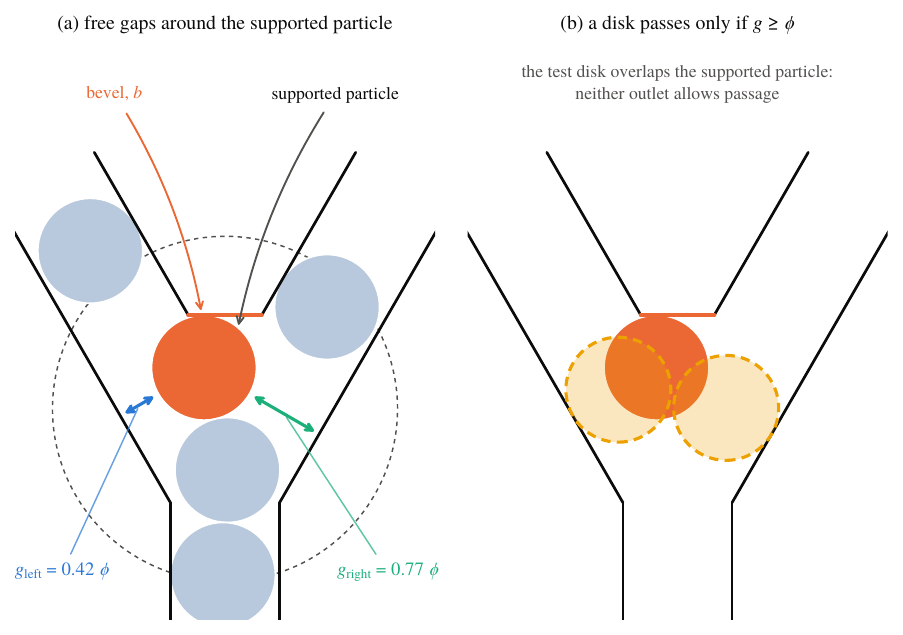}
	\caption{Geometrical constraints on particle passage near the truncated inner vertex of the Y-junction.}
	\label{figure5}
\end{figure}

Blocking configurations were characterized by the support position of the leading particle and the free gaps $g_{\mathrm{left}}$ and $g_{\mathrm{right}}$ between its surface and the outer walls (Fig.~\ref{figure5}). A gap smaller than $\phi$ prevents another disk from passing. Consequently, a single supported particle can obstruct both branch entrances, even asymmetrically, as illustrated in Fig.~\ref{figure5}.

Collective arch-like configurations were identified when at least two particles within the decision region either contacted the inner wedge (surface-to-wall gap below $0.02\phi$) or formed a side-by-side contacting pair with a longitudinal separation below $0.5\phi$. The first condition predominates, allowing two particles supported on the truncated vertex to block both entrances without necessarily touching each other.

\begin{figure}[H]
	\centering
	\includegraphics[width=1.0\textwidth,height=0.9\textheight,keepaspectratio]{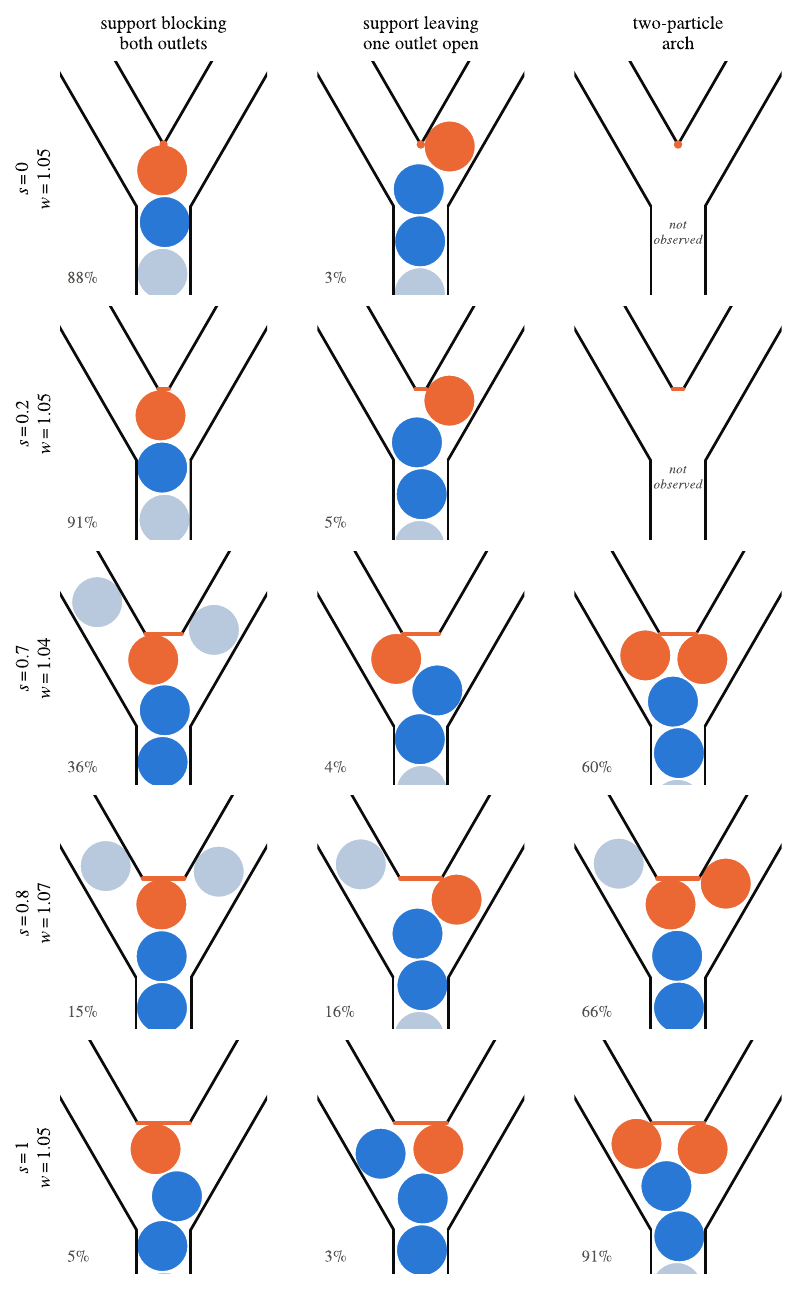}
	\caption{Representative blocking configurations for different truncations $s$: single-particle obstruction of both branches (left), single-particle support leaving one branch geometrically accessible (center), and collective two-particle arches (right). Percentages indicate the relative occurrence of each configuration for the corresponding $s$ and $w$.}
	\label{figure6}
\end{figure}

Figure~\ref{figure6} illustrates the main blocking mechanisms. Configurations were sampled periodically every $10^5$ MCS, during clogging episodes identified by $\chi(t)$, and when neither branch received a particle for more than $2\times10^6$ MCS. The last criterion captures persistent blockages involving fewer than four particles, which may escape detection by $\chi(t)$.

For the ideal vertex ($s=0$), a single supported particle may obstruct both branches but remains sensitive to transverse fluctuations. Small truncations provide more stable support, favoring persistent single-particle blocking and explaining the global maximum of $w_c$ near $s=0.2$.

At larger truncations ($s\geq0.7$), collective two-particle arches become the dominant blocking mechanism. Unlike persistent single-particle obstructions, these structures can rearrange and release the junction, consistent with the decrease of $w_c$ toward its minimum at $s=0.7$.

The secondary maximum at $s=0.8$ appears to reflect the coexistence of collective arches and persistent single-particle support. At this truncation, the flat vertex can accommodate laterally displaced supported particles, whereas at $s=1$ collective configurations become more prevalent. Because particle overtaking is forbidden in the narrow channels, a geometrically accessible branch does not necessarily allow the following particle to bypass a supported disk. This competition between blocking mechanisms may account for the local increase in $w_c$, although the observed configurations do not establish a quantitative causal relationship.

Overall, the non-monotonic dependence of $w_c(s)$ is consistent with a geometry-induced change in the dominant blocking mechanism: from persistent single-particle obstruction at small truncations to collective, dynamically reversible arches at larger truncations.

\subsection{Residence-time statistics and particle-order rearrangements}

\begin{figure}[htbp]
	\centering
	\includegraphics[width=0.78\linewidth]{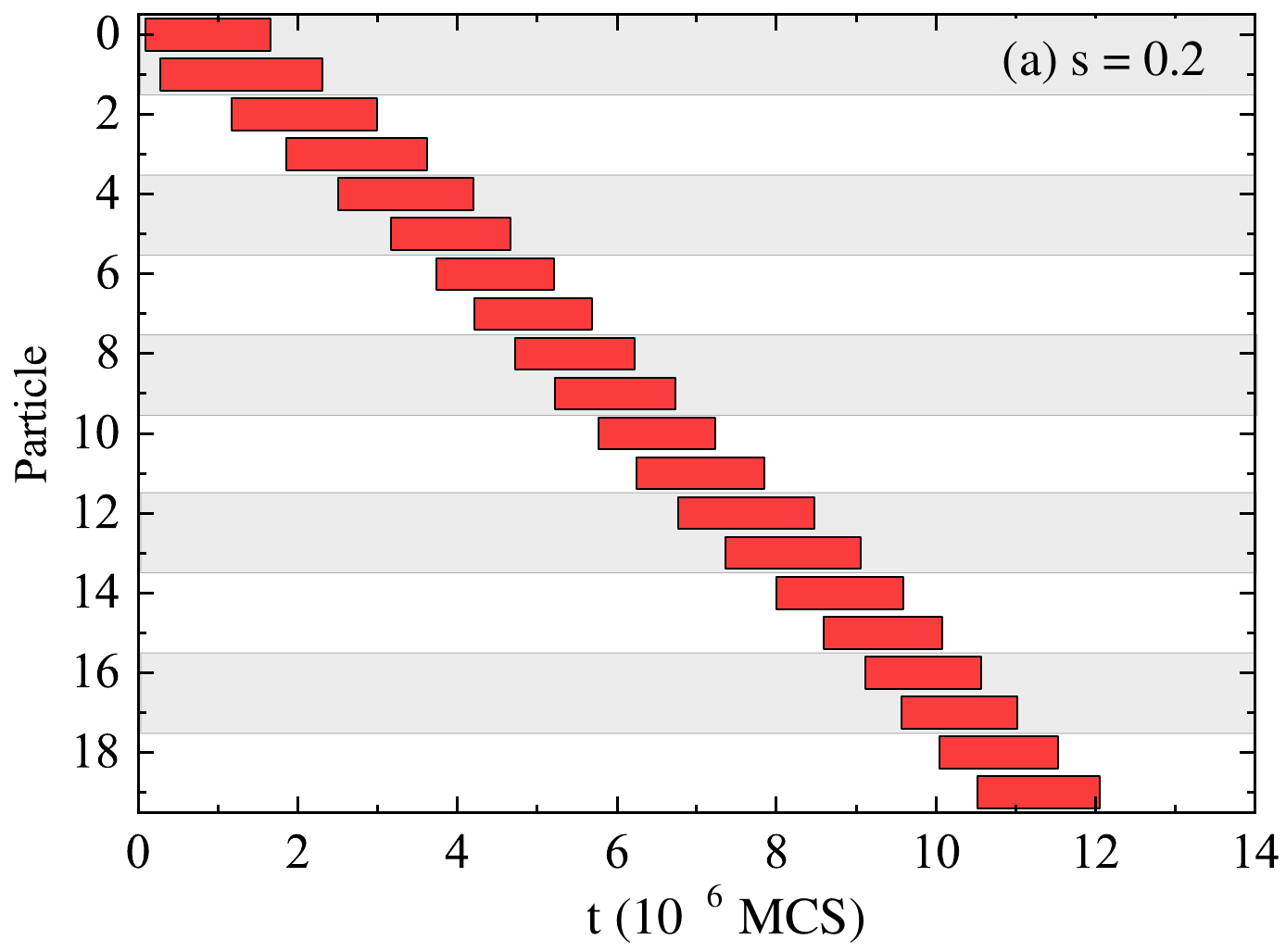}
	\vspace{0.15cm}
	\includegraphics[width=0.78\linewidth]{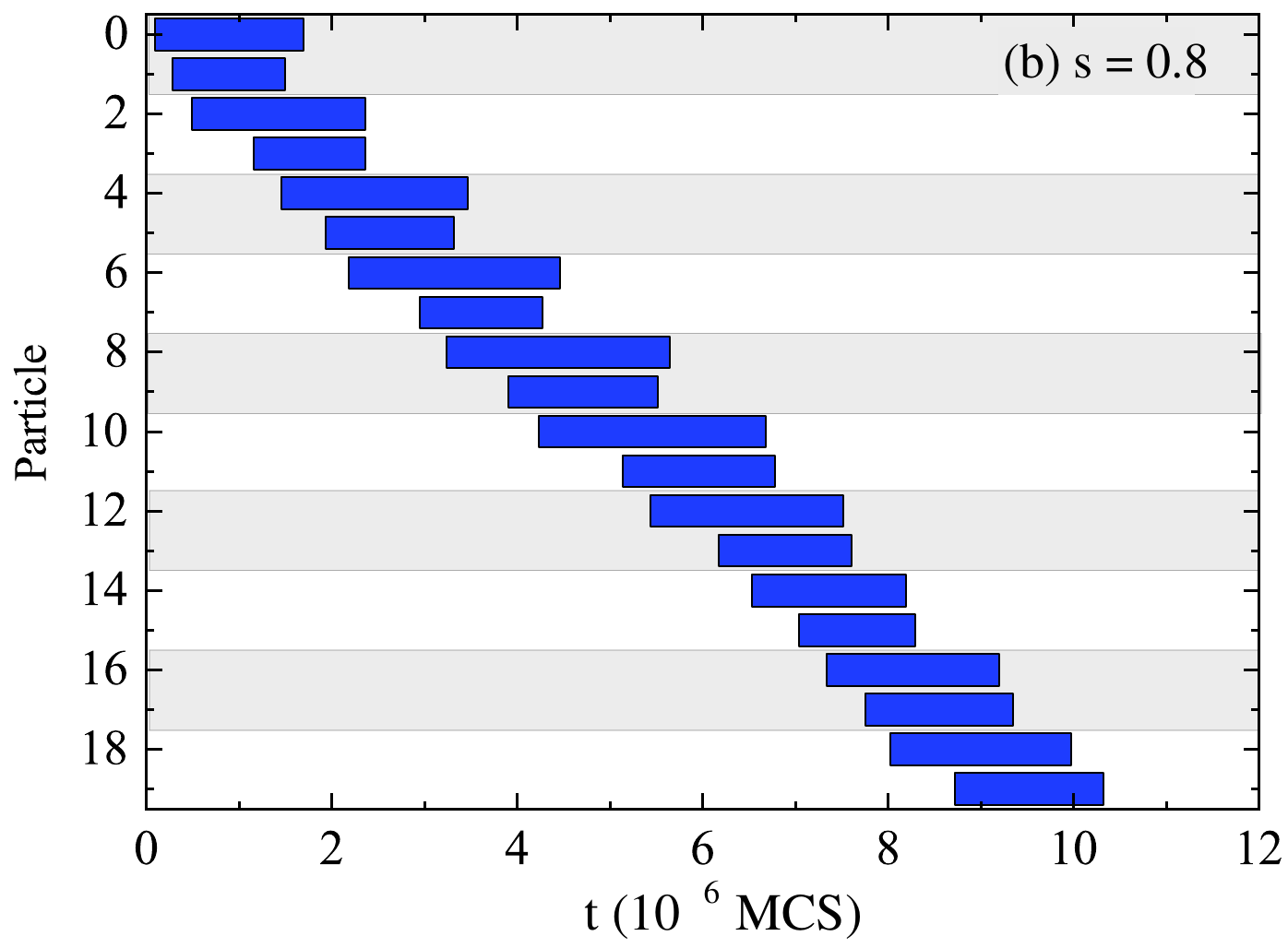}
	\caption{Residence intervals of successive
		particles for $w=1.09$: (a) $s=0.2$
		and (b) $s=0.8$. Horizontal bars indicate
		the time spent by each particle in the
		decision region, ordered by its initial
		position in the incoming channel.
		Sequential transport at $s=0.2$ contrasts
		with the overlapping residence intervals
		at $s=0.8$.}
	\label{fig:residence_intervals}
\end{figure}

The residence-time intervals reveal a geometry-dependent change in particle transport through the junction. As shown in Fig.~\ref{fig:residence_intervals}, particles traverse the decision region sequentially for $s=0.2$, preserving their entrance order. In contrast, for $s=0.8$, consecutive particles frequently occupy the region simultaneously and leave nearly together, indicating the emergence of collective rearrangements.

This change is quantified by the inversion fraction $f_{\mathrm{inv}}$. For $w=1.09$, particle ordering is preserved for $s\leq0.5$, whereas larger truncations allow order inversions. The inversion fraction reaches a maximum of $0.543\pm0.018$ at $s=0.8$, consistent with nearly uncorrelated entrance and exit ordering. For $s\geq0.9$, it decreases to approximately $0.15$--$0.21$. Since overtaking is forbidden in the straight channels, these rearrangements originate within the enlarged decision region.

Collective motion also produces differences between the residence times of consecutive particles. When two particles occupy the junction simultaneously and leave nearly together, the first generally remains longer than the second. The ratio of their mean excess residence times reaches $2.38$ at $s=0.7$ and $2.07$ at $s=0.8$, compared with values close to unity for $s\leq0.5$.

\begin{figure}[p]
	\centering
	\includegraphics[
	width=\textwidth,
	height=0.85\textheight,
	keepaspectratio
	]{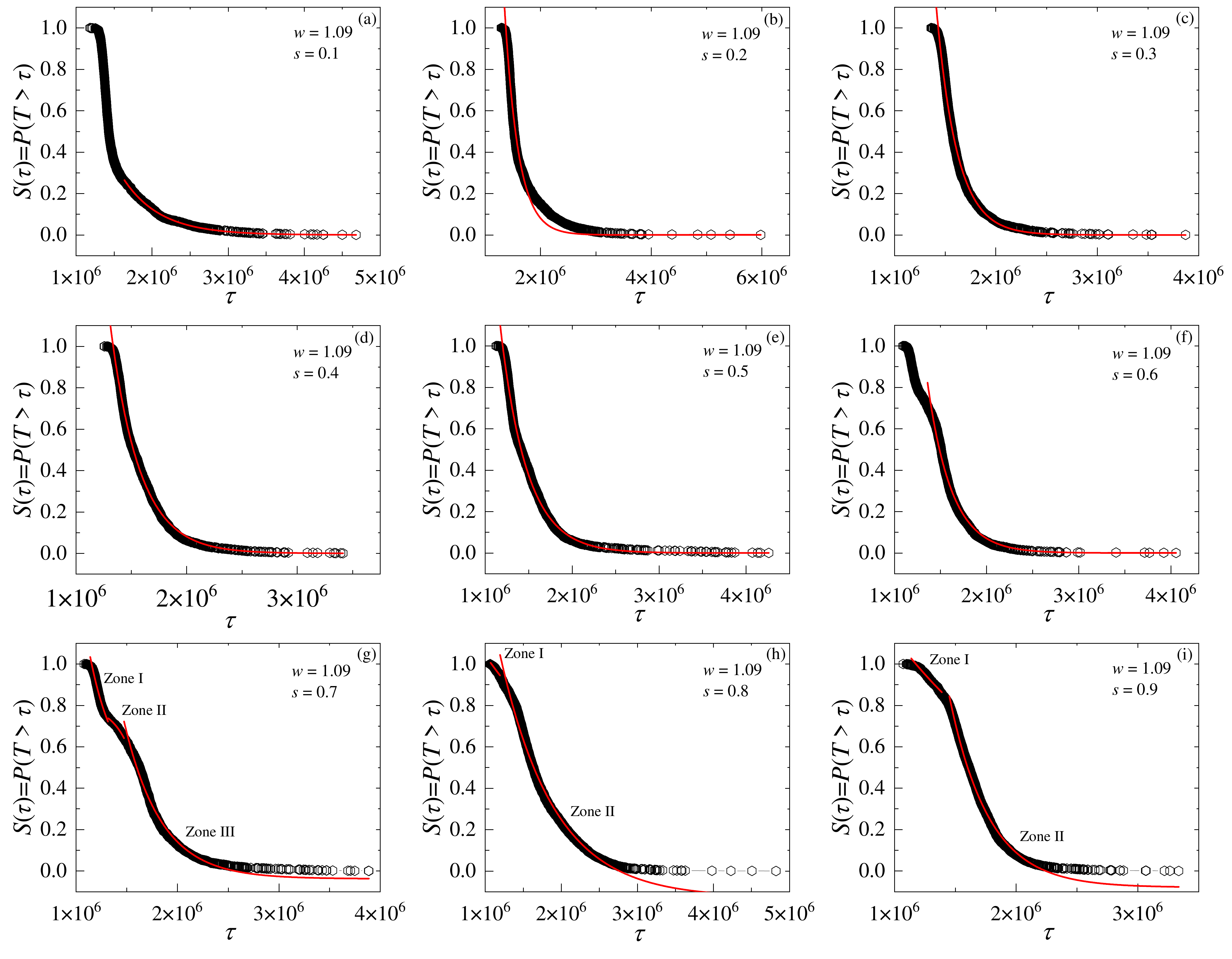}
	\caption{Residence-time survival functions $S(\tau)=P(T>\tau)$
		for $w=1.09$ and different truncation parameters:
		(a) $s=0.1$, (b) $s=0.2$, (c) $s=0.3$,
		(d) $s=0.4$, (e) $s=0.5$, (f) $s=0.6$,
		(g) $s=0.7$, (h) $s=0.8$, and (i) $s=0.9$.
		Black symbols represent simulation results, while red
		solid lines indicate exponential fits over selected
		time intervals. For small truncations, the survival
		functions exhibit an initial rapid decay followed by
		slower relaxation. As the truncation increases,
		the decay becomes progressively more complex,
		with distinct relaxation regimes identified in
		panels (g)--(i), revealing deviations from
		single-exponential behavior.}
	\label{fig:survival}
\end{figure}

Figure~\ref{fig:survival} shows the residence-time survival functions $S(\tau)$ for $w=1.09$. At small truncations, the curves exhibit relatively smooth decays, whereas for $s\geq0.7$ their nonexponential character becomes more pronounced. Local exponential fits describe selected intervals of the curves but do not imply distinct microscopic relaxation mechanisms.

For a more quantitative characterization, we consider the excess residence time $\tau-\tau_0$, where $\tau_0$ is the minimum observed residence time for each geometry. For $s\geq0.6$, its coefficient of variation ranges from $0.43$ to $0.77$, below the exponential value of unity. This behavior is incompatible with a mixture of unshifted exponential waiting times and is consistent with a sequential-stage description, in which the excess residence time results from the sum of several stochastic contributions. Such a description provides a possible interpretation of the nonexponential survival curves without assigning separate physical mechanisms to the fitted intervals.

\begin{figure}[htbp]
	\centering
	\includegraphics[width=0.5\linewidth]{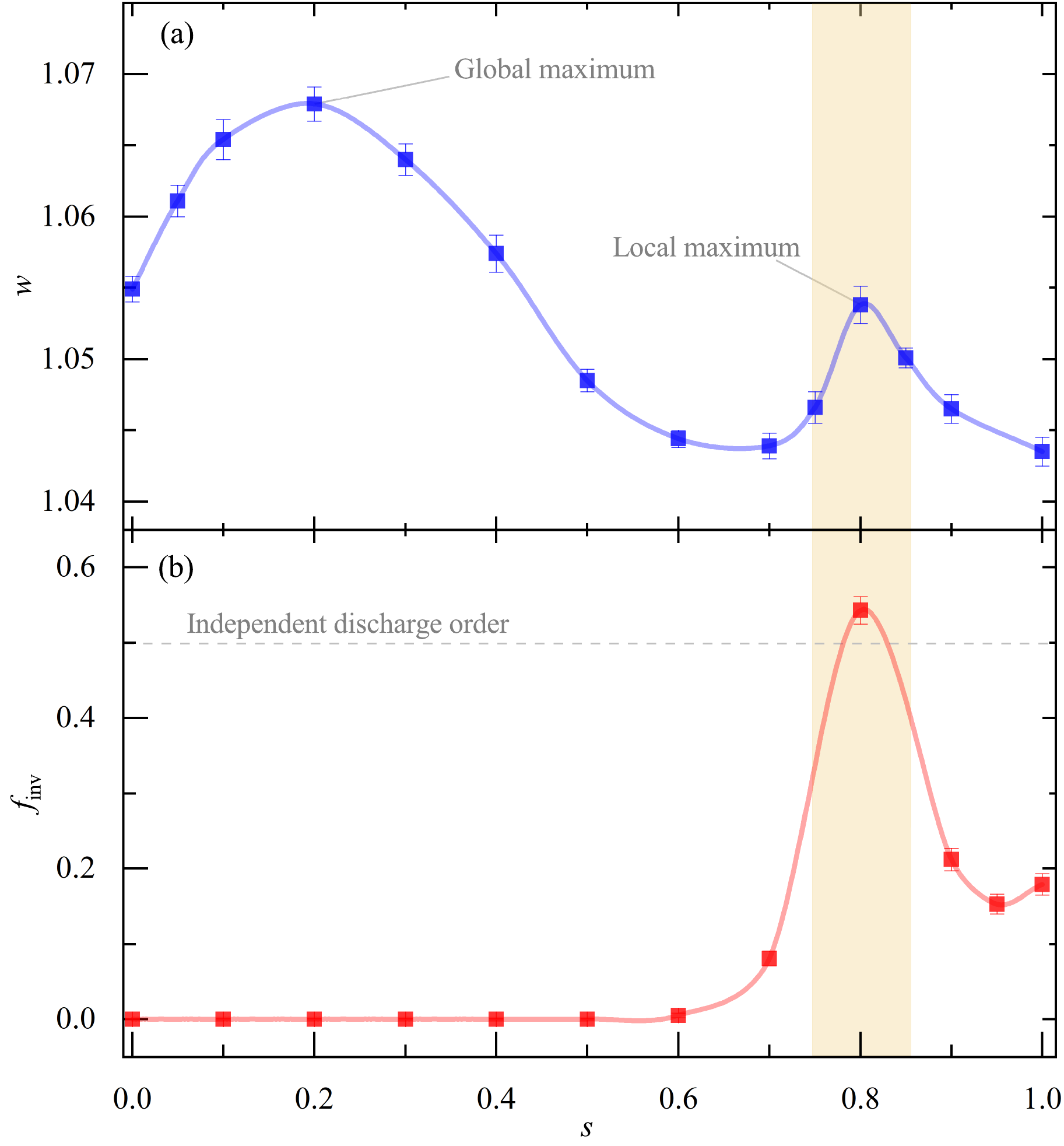}
	\caption{Characteristic width $w_c$ and inversion
		fraction $f_{\mathrm{inv}}$ as functions of $s$.
		The global maximum of $w_c$ occurs in the
		ordered regime ($s=0.2$), whereas its
		secondary maximum coincides with the
		maximum inversion fraction near $s=0.8$.}
	\label{fig:wc_finv}
\end{figure}

A particularly relevant connection emerges when $f_{\mathrm{inv}}(s)$ is compared with $w_c(s)$ (Fig.~\ref{fig:wc_finv}). Both quantities exhibit a local maximum at $s=0.8$, whereas the global maximum of $w_c$ at $s=0.2$ occurs in the strictly ordered regime. This distinction supports the existence of two geometrically controlled blocking mechanisms: persistent single-particle trapping at small truncations and collective rearrangements at larger truncations. The inversion fraction thus provides a complementary dynamical signature of the mechanisms underlying the non-monotonic dependence of $w_c(s)$.

\FloatBarrier

\section{Conclusions}
\label{sec:conclusions}

We have investigated how the local geometry of a confined Y-junction controls particle transport and clogging in a minimal system of hard disks interacting exclusively through excluded volume. Our results show that modifying the shape of the inner vertex, while keeping the channel width and branching angle fixed, produces a pronounced non-monotonic variation of the characteristic width \(w_c\), operationally defined by \(P_f(w_c)=1/2\). This demonstrates that the onset of sustained transport depends not only on the degree of confinement but also on the detailed morphology of the decision region.

The non-monotonic behavior of \(w_c(s)\) can be interpreted in terms of competing microscopic blocking mechanisms. Small truncations favor stable single-particle configurations supported by the inner vertex, consistent with the global maximum of \(w_c\) at \(s=0.2\). Larger truncations provide additional space for collective rearrangements and promote two-particle arch-like configurations. The competition between these mechanisms offers a physical interpretation of the subsequent decrease in \(w_c\) and the emergence of a secondary maximum near \(s=0.8\).

The dynamical analysis provides complementary evidence of this geometrically induced change in transport mechanisms. Particle-order inversions are absent for \(s\leq 0.5\), whereas larger truncations allow collective rearrangements within the decision region. Remarkably, the maximum inversion fraction at \(s=0.8\) coincides with the secondary maximum of \(w_c(s)\), while the global maximum occurs in the strictly ordered regime. Residence-time statistics further reveal deviations from simple exponential relaxation, compatible with a sequential-stage description of the excess residence times. Together, these observations indicate that the two maxima of \(w_c(s)\) are associated with distinct microscopic transport conditions.

Overall, our findings establish local junction geometry as an effective control parameter for clogging and particle reorganization in confined excluded-volume systems. Although the characteristic widths obtained here are finite-time operational thresholds, the identified mechanisms suggest experimentally testable strategies for controlling particle transport in microfluidic Y-junctions by modifying the vertex morphology without changing the channel width or branching angle.

\section*{CRediT authorship contribution statement}

Horacio Javier Diaz Carrera:
Methodology, Software, Investigation,
Data curation, Visualization,
Writing -- original draft.

Paulo M. Centres:
Conceptualization, Methodology, Software,
Supervision, Formal analysis,
Writing -- original draft,
Writing -- review \& editing.

Sergio J. Manzi:
Conceptualization, Supervision,
Writing -- review \& editing.

Diego J. Pérez-Morelo:
Methodology, Investigation,
Writing -- review \& editing.

\section*{Declaration of competing interest}
The authors declare that they have no known competing financial interests or personal relationships that could have appeared to influence the work reported in this paper.

\section*{Data availability}

The data and simulation code supporting the findings of this study will be made available upon reasonable request.

\section*{Acknowledgments}

The authors acknowledge financial support from the Consejo Nacional de Investigaciones Científicas y Técnicas (CONICET, Argentina) through PIP project KE 11220220100321CO, the Universidad Nacional de San Luis (UNSL) through PROIPRO project 03-2523, and Fundación Williams.

The authors also express their gratitude to Dr. Jorge Ernesto Castro, Dean of the Facultad de Ciencias Exactas, Físicas y Naturales at the Universidad Nacional de San Juan, for his financial support, and to Dr. Víctor D. Pereyra and Dr. Diego Ozcoidi Maza for their valuable contributions and insightful comments.

Finally, the authors acknowledge the fundamental role of free, publicly funded universities in Argentina in supporting scientific research, higher education, and the advancement of knowledge.

\appendix

\section{Sensitivity to simulation parameters}
\setcounter{figure}{0}
\renewcommand{\thefigure}{A.\arabic{figure}}
\subsection{Sensitivity to the transverse Monte Carlo step}
\label{app:step_size}

To assess the sensitivity of the characteristic width $w_c(s)$ to the transverse Monte Carlo step, additional simulations were performed using $n_T=3000$, while the main results were obtained with $n_T=2000$. In both cases, the transverse displacement is defined as
\begin{equation}
	\Delta x=\frac{(w-1)\phi}{n_T}.
\end{equation}

\begin{figure}[p]
	\centering
	\includegraphics[
	width=0.95\textwidth,
	keepaspectratio
	]{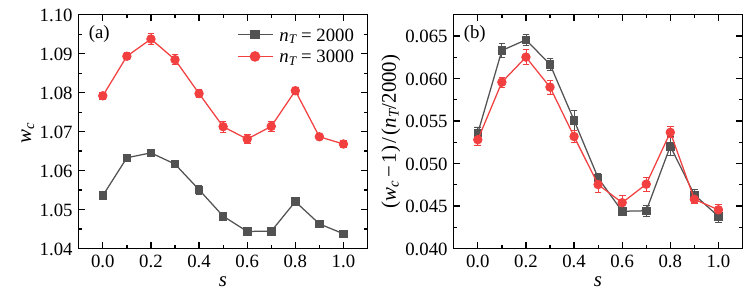}
	
	\caption{Comparison of the characteristic widths
		obtained with $n_T=2000$ and $n_T=3000$.
		The normalized representation highlights
		the preservation of the non-monotonic dependence
		on the truncation parameter $s$.}
	
	\label{fig:appendix_step}
\end{figure}

As shown in Fig.~\ref{fig:appendix_step}(a),
the absolute values of $w_c$ depend on the
transverse step size. However, the non-monotonic
dependence on the truncation parameter $s$,
including the global and secondary maxima,
is preserved for both choices of $n_T$.

The relative displacement of the characteristic widths satisfies approximately
\begin{equation}
	\frac{w_c(s;3000)-1}{w_c(s;2000)-1}
	=1.48\pm0.02.
\end{equation}

This ratio is close to $3000/2000=1.5$, indicating an approximately proportional scaling of the excess characteristic width with $n_T$ over the investigated range.

When the characteristic widths are normalized
by their corresponding transverse-step factors,
the resulting curves exhibit an approximate
collapse, with a mean relative difference
of about 3\%, as shown in
Fig.~\ref{fig:appendix_step}(b).

These results indicate that the non-monotonic dependence of $w_c(s)$ is robust against the tested variation of the transverse Monte Carlo step. Nevertheless, the absolute characteristic widths remain dependent on the simulation dynamics, and the present comparison does not establish convergence in the limit of vanishing step size or infinite observation time.

\subsection{Sensitivity to the observation time}
\label{app:observation_time}

To assess the influence of the finite observation time on
the characteristic flow width, additional simulations
were performed using observation windows of
$T_{\mathrm{obs}}=3\times10^{7}$ and $4\times10^{7}$ MCS,
in addition to the reference value
$T_{\mathrm{obs}}=2\times10^{7}$ MCS used throughout
the main text.

\begin{figure}[htbp]
	\centering
	\includegraphics[width=0.60\columnwidth]{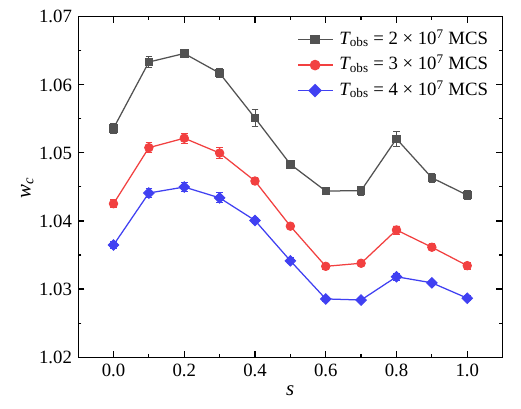}
	\caption{
		Dependence of the characteristic flow width $w_c$
		on the junction truncation parameter $s$ for
		different observation times:
		$T_{\mathrm{obs}}=2\times10^{7}$,
		$3\times10^{7}$, and $4\times10^{7}$ MCS.
		Increasing the observation time shifts the
		characteristic widths toward smaller values,
		while preserving the non-monotonic dependence
		on the junction geometry.
	}
	\label{fig:time_sensitivity}
\end{figure}

Figure~\ref{fig:time_sensitivity} shows the resulting
dependence of $w_c$ on the junction truncation parameter
$s$. Increasing the observation time systematically shifts
the characteristic widths toward smaller values.
This behavior is consistent with the finite-time
definition of the flow probability: longer observation
windows allow additional particle transits to be recorded.

Importantly, the qualitative structure of $w_c(s)$
remains unchanged over the investigated observation
times. In particular, the main maximum near $s=0.2$,
the subsequent decrease toward $s\approx0.6$--$0.7$,
and the secondary maximum near $s=0.8$ persist
across all three observation windows.

These results indicate that the non-monotonic dependence
of the characteristic flow width on the junction geometry
is robust against the investigated variation of the
observation time, although the absolute values of $w_c$
remain time dependent.

Together with the transverse-step sensitivity analysis,
this provides additional evidence that the observed
geometrical trends are not restricted to a single choice
of simulation parameters. Nevertheless, the present
results do not establish convergence in the limit
$T_{\mathrm{obs}}\rightarrow\infty$.

\FloatBarrier


\begin{thebibliography}{00}
	
	\bibitem{Hodgkin}
	A. L. Hodgkin and R. D. Keynes,
	J. Physiol. \textbf{128}, 61 (1955).
	
	\bibitem{Taloni}
	A. Taloni, O. Flomenbom, R. Castañeda-Priego,
	F. Marchesoni,
	Soft Matter \textbf{13}, 1096 (2017);
	and references therein.
	
	\bibitem{Harris}
	T. Harris,
	Journal of Applied Probability \textbf{2}, 323 (1965).
	
	\bibitem{Jepsen}
	D. W. Jepsen,
	J. Math. Phys. \textbf{6}, 405 (1965).
	
	\bibitem{Kasperkovitz}
	P. Kasperkovitz and J. Reisenberger,
	Phys. Rev. A \textbf{31}, 2639 (1985).
	
	\bibitem{Arriata}
	R. Arratia,
	Ann. Probab. \textbf{11}, 362--373 (1983).
	
	\bibitem{Richards}
	P. M. Richards,
	Phys. Rev. B \textbf{16}, 1393 (1977).
	
	\bibitem{Fedders}
	P. A. Fedders,
	Phys. Rev. B \textbf{17}, 40 (1978).
	
	\bibitem{Alexander}
	S. Alexander and P. Pincus,
	Phys. Rev. B \textbf{18}, 2011 (1978).
	
	\bibitem{Lizana}
	L. Lizana and T. Ambj\"ornsson,
	Phys. Rev. Lett. \textbf{100}, 200601 (2008).
	
	\bibitem{Beijeren}
	H. van Beijeren, K. W. Kehr, and R. Kutner,
	Phys. Rev. B \textbf{28}, 5711 (1983).
	
	\bibitem{Coste}
	C. Coste, J.-B. Delfau, C. Even, and M. Saint Jean,
	Phys. Rev. E \textbf{81}, 051201 (2010).
	
	\bibitem{Delfau}
	J.-B. Delfau, C. Coste, and M. Saint Jean,
	Phys. Rev. E \textbf{84}, 011101 (2011).
	
	\bibitem{Tkachenko}
	D. V. Tkachenko, V. R. Misko, and F. M. Peeters,
	Phys. Rev. E \textbf{82}, 051102 (2010).
	
	\bibitem{Centres3}
	P. M. Centres, V. D. Pereyra, S. J. Manzi,
	and S. Bustingorry,
	Physica A \textbf{629}, 129225 (2023).
	
	\bibitem{Kutner}
	R. Kutner, H. van Beijeren, and K. W. Kehr,
	Phys. Rev. B \textbf{30}, 4382 (1984).
	
	\bibitem{Lucena}
	D. Lucena, D. V. Tkachenko, K. Nelissen,
	V. R. Misko, W. P. Ferreira, G. A. Farias,
	and F. M. Peeters,
	Phys. Rev. E \textbf{85}, 031147 (2012).
	
	\bibitem{Mon}
	K. K. Mon and J. K. Percus,
	J. Chem. Phys. \textbf{112}, 3457 (2000).
	
	\bibitem{Jimaan}
	J. San\'e, J. T. Padding, and A. A. Louis,
	Phys. Rev. E \textbf{79}, 051402 (2009).
	
	\bibitem{Ooshida}
	Takeshi Ooshida  and Michio Otsuki,
	Journal of the Physical Society of Japan
	\textbf{86}, 113002 (2017).
	
	\bibitem{Centres4}
	S. J. Manzi, S. Bustingorry, and P. M. Centres,
	Physica A \textbf{675}, 130808 (2025).
	
	\bibitem{Larimi}
	M. M. Larimi, A. Ramiar, and A. A. Ranjbar,
	Journal of Magnetism and Magnetic Materials
	\textbf{362}, 58--71 (2014).
	
	\bibitem{Antonov2025}
	A. P. Antonov, M. Terkel, F. J. Schwarzendahl,
	C. Rodríguez-Gallo, P. Tierno, and H. Löwen,
	Communications Physics \textbf{8}, 165 (2025).
	
	\bibitem{Bushi2005}
	D. Bushi, Y. Grad, S. Einav, O. Yodfat,
	B. Nishri, and D. Tanne,
	Stroke \textbf{36}, 2696--2700 (2005).
	
	\bibitem{Garstecki2006}
	P. Garstecki, M. J. Fuerstman, H. A. Stone,
	and G. M. Whitesides,
	Lab on a Chip \textbf{6}, 437 (2006).
	
	\bibitem{Zhang2014}
	J. Zhang, S. Yan, R. Sluyter, W. Li, G. Alici,
	and N.-T. Nguyen,
	Scientific Reports \textbf{4}, 4527 (2014).
	
	\bibitem{Zuriguel2005}
	I. Zuriguel Ballaz,
	Ph.D. Thesis, Universidad de Navarra (2005).
	
	\bibitem{To2001}
	K. To, P.-Y. Lai, and H. K. Pak,
	Physical Review Letters \textbf{86}, 71 (2001).
	
	\bibitem{To2002}
	K. To, P.-Y. Lai, and H. K. Pak,
	Physica A \textbf{315}, 174 (2002).
	
	\bibitem{To2005}
	K. To,
	Physical Review E \textbf{71}, 060301 (2005).
	
	\bibitem{Hong2017}
	X. Hong, M. Kohne, M. Morrell, H. Wang,
	and E. R. Weeks,
	Phys. Rev. E \textbf{96}, 062605 (2017).
	
\end{thebibliography}
\end{document}